\documentclass[
reprint,
amsmath,
amssymb,
prl,
superscriptaddress,
notitlepage
]{revtex4-1}

\usepackage[colorlinks, breaklinks=true, linkcolor=blue, citecolor=blue, linktocpage=true]{hyperref}
\usepackage{color}
\usepackage{graphicx}
\usepackage{dcolumn}
\usepackage{bm}
\usepackage{ulem}

\usepackage{txfonts}

\begin{document}

\title{Tunable Charged Domain Wall from Topological Confinement\\in Nodal-Line Semimetals}

\author{Akihiko Sekine}
\email{akihiko.sekine@riken.jp}
\affiliation{RIKEN Center for Emergent Matter Science, Wako, Saitama 351-0198, Japan}
\author{Naoto Nagaosa}
\affiliation{RIKEN Center for Emergent Matter Science, Wako, Saitama 351-0198, Japan}
\affiliation{Department of Applied Physics, The University of Tokyo, Bunkyo, Tokyo 113-8656, Japan}

\date{\today}

\begin{abstract}
We study theoretically the electronic structure of topological nodal-line semimetals.
We show that, in the presence of a gap-opening spatially dependent mass term that forms a domain wall, an in-gap charged localized mode emerges at the domain wall.
It turns out that such a domain wall is realized by head-to-head (or tail-to-tail) bulk electric polarizations.
The localized mode has a topological origin, i.e., a topological confinement is realized, which is understood by a semiclassical topological number defined in the semiclassical momentum-real space.
In contrast to previous studies, our study demonstrates a topological confinement at the interface between two insulators without bulk topological numbers.
Moreover, the dispersion of the localized mode evolves from gapless to gapped as the bulk bandgap increases, which means that its conductivity is externally tunable.
We discuss a possible experimental realization of the stable, electrically-tunable charged domain wall.
\\
\end{abstract}

\maketitle

%%%%%%%%%%%%
{\it Introduction.---}The physics involving domain walls (DWs), where an order parameter of a system is spatially dependent and changes its sign, is one of the important research themes in modern physics.
From the viewpoint of device applications, the DWs separate domains that store information, which means that the creation and manipulation of DWs on demand are of practical interest.
The charged DW, originating from the normal component of two bulk electric polarizations, has also attracted attention recently in the pursuit of future nanoelectronic devices \cite{Seidel2009,Catalan2012,Meier2015,Bednyakov2018}.
In contrast to magnetic DWs in spintronics, which usually require the bulk to be metallic to utilize current-induced methods \cite{Parkin2008}, the charged DW can be realized in insulating materials, and has so far been reported mostly in ferroelectrics such as BaTiO$_3$ and multiferroics such as BiFeO$_3$ \cite{Seidel2009,Catalan2012,Meier2015,Bednyakov2018,Matsubara2015}.

Topological materials are considered a promising materials class for future nanotechnological applications due to their peculiar nature arising from the topology.
For example, possible ways to manipulate and utilize the spin-momentum locked surface states of topological insulators have been investigated experimentally in spintronics \cite{Mellnik2014,Li2014,Fan2014,Shiomi2014}.
The focus of this Letter is on a new class of three-dimensional (3D) topological semimetals, called the nodal-line semimetals (NLSMs), in which the conduction and valence bands touch along a line or loop.
Although several theoretical predictions \cite{Mullen2015,Kim2015,Yu2015,Fang2015,Xie2015,Yamakage2016,Chan2016,Ezawa2016,Hirayama2017} and experimental realizations \cite{Bian2016,Schoop2016,Hu2016,Takane2016,Wang2017,Liu2018,Takane2018,Lou2018} of NLSMs have been made so far, the electronic properties of NLSMs are not yet well understood.

In this Letter, we investigate the electronic structure of topological NLSMs in the presence of a gap-opening spatially dependent mass term that forms a DW.
We show that, in the case of the DW parallel to the nodal-line plane, the DW originates from head-to-head (or tail-to-tail) bulk electric polarizations, and that an in-gap charged localized mode which presence is topologically protected emerges at the DW.
Namely, a stable charged DW of electric polarizations is realized in NLSMs.
In sharp contrast to the well-known band bending mechanism \cite{Bednyakov2018}, the origin of the charged DW in this study is purely electronic, i.e., due to the band topology.
This study demonstrates the emergence of a topologically confined state at the interface between two insulators without bulk topological numbers, whereas it has been understood that topologically confined states emerge at the interface between two insulators as a result of the difference between their bulk topological numbers \cite{Jackiw1976,Su1979,Martin2008,Semenoff2008,Yao2009,Jung2011,Ju2015,Comment6}.
In addition, the dispersion of the localized mode evolves from gapless to gapped as the bulk bandgap increases, which means that its conductivity is externally tunable.

%%%%%%%%%%%%
%
\begin{figure}[!b]
\centering
\includegraphics[width=0.95\columnwidth]{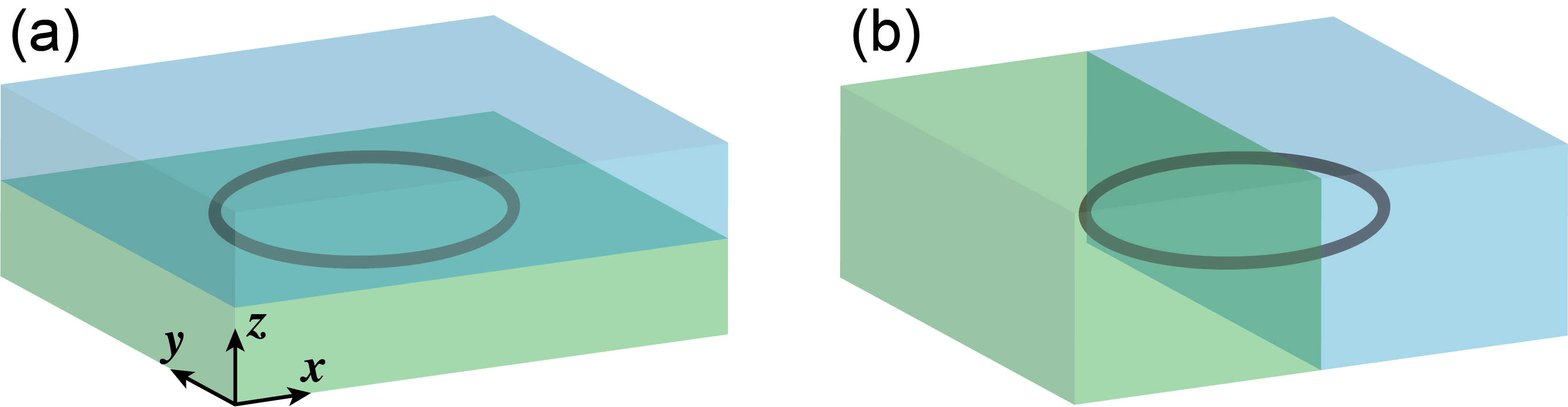}
\caption{Schematic illustration of a NLSM with a spatially dependent mass term (a) in the $z$ direction and (b) in the $x$ direction.
The nodal line is located on the $x$-$y$ plane.
Two colors (green and blue) indicate the sign of the mass $V$ that is different from each other.
}
\label{Fig1}
\end{figure}
\begin{figure*}[!t]
\centering
\includegraphics[width=2\columnwidth]{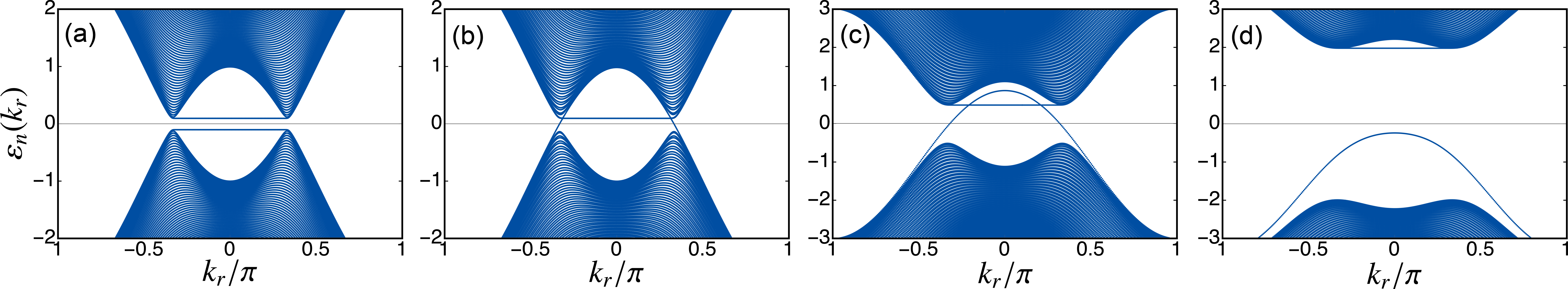}
\caption{Energy spectrum (a) in the presence of a uniform mass with $V_0=0.1$, and in the presence of a parallel DW with (b) $V_0=0.1$, (c) $V_0=0.5$, and (d) $V_0=2$.
In (a) [(b)] the dispersion of the localized mode around the zero-energy level is linear (nonlinear) in $k_r$.
In (a)-(d) we set $k_0=m_1=m_2=1$ and $v_z=2$.
Note that spectrum is inverted, i.e., $\varepsilon_n(k_r)\to -\varepsilon_n(k_r)$, under the operation $V_0\to -V_0$.
}
\label{Fig2}
\end{figure*}

{\it Model.---}We start from the minimal two-band model for $\mathcal{PT}$-symmetric NLSMs without spin-orbit coupling such as CaAgP and Ca$_3$P$_2$ \cite{Yamakage2016,Chan2016,Rui2018}
\begin{align}
\mathcal{H}(\bm{k})=\bigl[k_0^2-m_1(k_x^2+k_y^2)-m_2k_z^2\bigr]\sigma_x+v_z k_z\sigma_y+V\sigma_z,
\label{Hamiltonian-continuum}
\end{align}
where $\sigma_i$ are the Pauli matrices acting on an orbital space and we have introduced a gap-opening mass $V$ that breaks $\mathcal{PT}$-symmetry \cite{Comment7}.
Note that in Eq.~(\ref{Hamiltonian-continuum}) we have unitary transformed the original Hamiltonian obtained in Refs.~\cite{Yamakage2016,Chan2016,Rui2018} as $\sigma_z\to\sigma_x$ and $\sigma_x\to\sigma_z$ for later convenience.
When $V=0$, the system has a nodal line given by the circle $k_x^2+k_y^2=k_0^2/m_1$ with $k_z=0$, and is topological since the Zak phase (or equivalently the Berry phase in this case) takes a quantized value \cite{Comment5}.
When $V$ is finite and independent of spatial position, the energy eigenvalues of the Hamiltonian~(\ref{Hamiltonian-continuum}) are readily obtained as $E(\bm{k})=\pm\sqrt{[k_0^2-m_1(k_x^2+k_y^2)-m_2k_z^2]^2+v_z^2 k_z^2+V^2}$, which shows a fully gapped spectrum.

%%%%%%%%%%%%
{\it Tunable localized mode at the domain wall.---}Let us introduce two simplest configurations of the spatially dependent DW structure in the mass $V$, as schematically shown in Fig.~\ref{Fig1}.
We call a DW in the $z$ direction [i.e., $V(z)=V_0\mathrm{sgn}(z)$] the ``parallel DW'', and a DW in the $x$ direction [i.e., $V(x)=V_0\mathrm{sgn}(x)$] the ``perpendicular DW''.
To investigate the effect of the presence of the DW on the electronic structure of the system, we consider a lattice version of Eq.~(\ref{Hamiltonian-continuum}) in a slab geometry where the thickness is finite in the DW direction but is infinite in the other two directions \cite{Comment-SM}.

Let us consider the case of the parallel DW, $V(z)=V_0\mathrm{sgn}(z)$.
In this case, it is useful to define the momentum $k_r$ satisfying $k_r^2=k_x^2+k_y^2$, since the system has rotation symmetry on the $k_x$-$k_y$ plane.
For reference we show the spectrum for the uniform mass case [i.e., $V(z)=V_0$] in Fig.~\ref{Fig2}(a), from which we see that the energy gap of $2|V_0|$ opens.
The two flat bands between the bulk band extrema $k_r=\pm k_{\mathrm{NL}}$ are the surface states well known in NLSMs \cite{Mullen2015,Kim2015,Yu2015,Fang2015,Yamakage2016,Chan2016}.
In the presence of the DW, we find an in-gap localized mode at $z=0$, as shown in Figs.~\ref{Fig2}(b)-(d).
Note that there still exist the surface states at the top and bottom surfaces, and their flat dispersions are doubly degenerate at $\varepsilon=V_0$.
We numerically find that, when the DW mass $V_0$ is small, the localized mode around $k_r=\pm k_{\mathrm{NL}}$ has a linear dispersion as
\begin{align}
\varepsilon_0(k_r)=\mp v_F \bigl(k_r \mp k_r^0\bigr),
\end{align}
as shown in Figs.~\ref{Fig2}(b).
As the mass $V_0$ becomes larger, the dispersion of the localized mode becomes nonlinear in $k_r$ and eventually becomes gapped, i.e., fails to cross the zero-energy level (the $\varepsilon=0$ line), as shown in Figs.~\ref{Fig2}(c) and \ref{Fig2}(d).

We also find the existence of a localized mode at the perpendicular DW, $V(x)=V_0\mathrm{sgn}(x)$ \cite{Comment-SM}.
However, the dispersion of the localized mode appearing at the perpendicular DW is always gapped regardless of the value of $V_0$, which implies that its origin is not topological as shall be shown later.
Therefore, in what follows we focus on the parallel DW case.

%%%%%%%%%%%%
{\it Semi-analytical solution for the localized modes.---}So far we have seen the presence of two kinds of the localized modes at the DWs, depending on the direction of the DW.
Here, we confirm the presence of such localized modes by a different method, starting from the continuum model (\ref{Hamiltonian-continuum}).
To introduce the spatial dependence in the $z$ direction, we need to set $k_z\to -i\partial_z$ in Eq.~(\ref{Hamiltonian-continuum}). Then, the Schroedinger equation $\mathcal{H}(k_r, z)\Psi(z)=\varepsilon\Psi(z)$ with $\Psi(z)=[u(z),v(z)]^T$ reads
\begin{subequations}
\begin{align}
V(z)u+\bigl[k_0^2-m_1 k_r^2+m_2\partial_z^2\bigr]v-v_z\partial_z v =\varepsilon u,
\label{First-Eq}
\\
\nonumber\\
-V(z)v+\bigl[k_0^2-m_1 k_r^2+m_2\partial_z^2\bigr]u+v_z\partial_z u =\varepsilon v.
\label{Second-Eq}
\end{align}
\end{subequations}
It can be easily checked that, when $V(z)$ has a generic DW structure satisfying $V(-z)=-V(z)$, Eq.~(\ref{Second-Eq}) with $v(z)=u(-z)$ or $v(z)=-u(-z)$ is identical to Eq.~(\ref{First-Eq}).
Namely, the solution of the wave functions for the Schroedinger equation has the two forms: $\Psi(z)=[u(z),u(-z)]^T$ or $\Phi(z)=[u(z),-u(-z)]^T$.

We consider the simplest DW structure, $V(z)=V_0\mathrm{sgn}(z)$, as it has been considered similarly for bilayer graphene \cite{Martin2008}.
Since the wave function $\Psi(z)$ of the mode localized at $z=0$ should be constant in the region where $z$ is far away from zero, we can assume without loss of generality the solution of the form $\Psi(z)\propto e^{-\lambda z}$.
For concreteness let us consider the case of $\Psi(z)=[u(z),u(-z)]^T$ \cite{Comment3}.
Then, we have $u(z)=u_0e^{-\lambda z}$ and $v(z)=u(-z)=u_0e^{\lambda z}$.
Substituting these into Eqs.~(\ref{First-Eq}) and (\ref{Second-Eq}) and eliminating $e^{\lambda z}$, we obtain the solution for $\lambda$:
\begin{align}
\lambda=\frac{v_z\pm \sqrt{v_z^2-4m_2\left[k_0^2-m_1 k_r^2\mp i\sqrt{V_0^2-\varepsilon^2}\right]}}{2m_2},
\label{Solution-lambda}
\end{align}
for the in-gap states with $|\varepsilon|<V_0$.
We see from Eq.~(\ref{Solution-lambda}) that the real part of $\lambda$ can be positive or negative.
Then, the general form of $u(z)$ takes a different form depending on the region with $z>0$ or $z<0$ as follows:
\begin{align}
u^+(z)&=u_1^+e^{-\lambda_1^+ z}+u_2^+e^{-\lambda_2^+ z}\ \ \ (\mathrm{for\ }z>0),\nonumber \\
u^-(z)&=u_1^-e^{-\lambda_1^- z}+u_2^-e^{-\lambda_2^- z}\ \ \ (\mathrm{for\ }z<0),
\end{align}
where the real part of $\lambda_{1,2}^+$ ($\lambda_{1,2}^-$) is positive (negative).
Since there are four unknown constants $u_{1,2}^{\pm}$, we need four boundary conditions for $u^{\pm}(z)$, which can be combined into a $4\times 4$ matrix form.
Finally, we obtain the energy spectrum of the localized mode $\varepsilon(k_r)$ from the condition such that the determinant of the $4\times 4$ matrix is zero \cite{Comment-SM}.
A Similar calculation can be applied to the case of the perpendicular DW \cite{Comment-SM}.

Although it is difficult to obtain the analytical expression for the energy spectrum $\varepsilon(k_r)$ as a function of arbitrary parameters, it is possible to obtain an analytical expression for the momentum of the zero-energy state satisfying $\varepsilon(k_r^0)=0$ with some parameters fixed (for example, $k_0=m_1=m_2=1$ here) as
\begin{align}
k_r^0=\sqrt{1-\frac{v_z^2}{4}-\frac{V_0^2 v_z^2}{\sqrt[3]{2}F(V_0, v_z)} + \frac{F(V_0, v_z)}{\sqrt[3]{2^{5}} v_z^2}},
\label{wavenumber-analytical}
\end{align}
where $F(V_0, v_z)=[4 V_0^2 v_z^8-V_0^4 v_z^4+(16 V_0^4 v_z^{16}+8 V_0^6 v_z^{12}+V_0^8 v_z^8)^{1/2}]^{1/3}$.
We find that the obtained results of the present continuum model are in qualitative agreement with the lattice model considered before in the behavior such that the localized mode crosses the zero-energy level and becomes fully gapped as the value of $V_0$ becomes larger \cite{Comment-SM}.
We also numerically confirm the change in the $k_r$ dependence of $\varepsilon(k_r)$ around zero energy from linear to nonlinear as the value of $V_0$ becomes larger, which has been seen in the lattice model.

\begin{figure}[!t]
\centering
\includegraphics[width=\columnwidth]{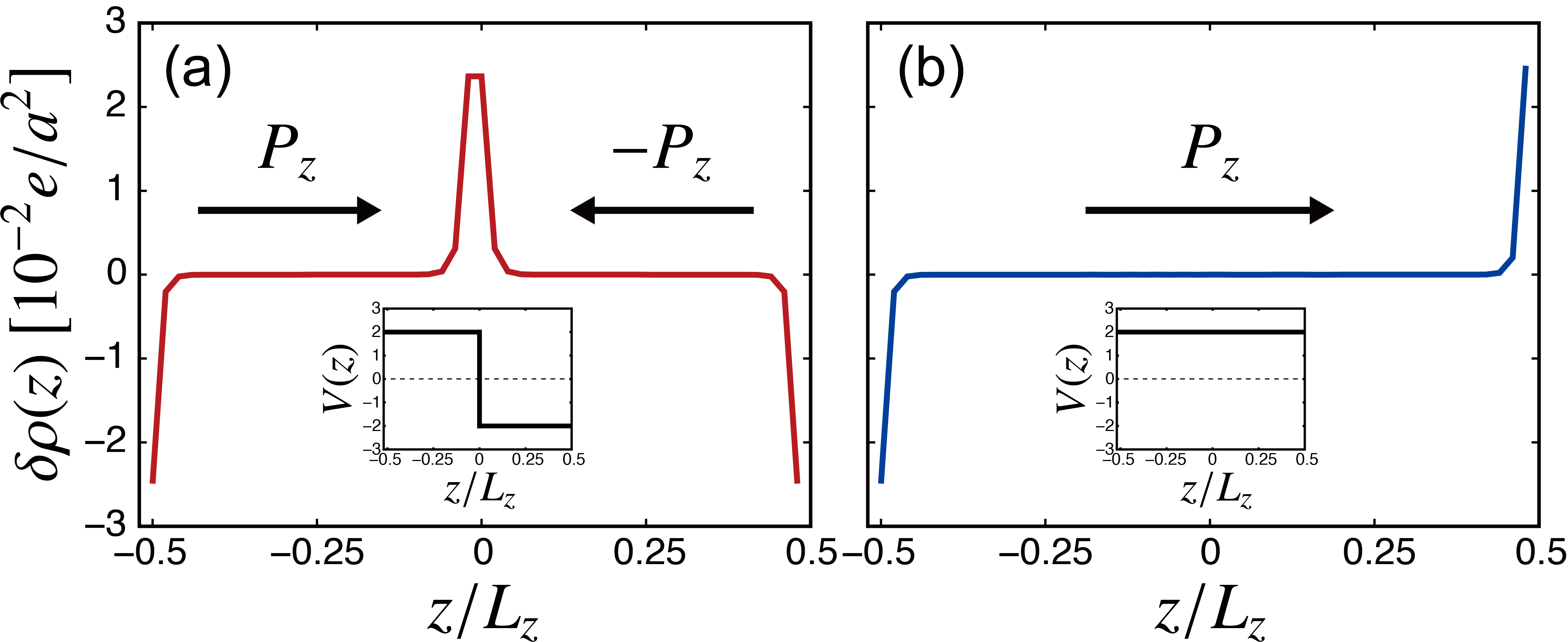}
\caption{$z$ dependence of the charge density $\delta\rho(z)$ (a) in the presence of a parallel DW, and (b) in the presence of a uniform mass.
The insets show the $z$ dependences of $V(z)$.
In (a) and (b) we set $k_0=m_1=m_2=1$, $v_z=2$, and $V_0=2$.
In both cases $\delta\rho(z)$ is inverted under the operation $V_0\to -V_0$.
}
\label{Fig3}
\end{figure}

%%%%%%%%%%%%
{\it Charges of the localized modes.---}In order to characterize the localized modes, we calculate the layer-resolved charge density for the lattice model with the parallel DW [$V(z)=V_0\mathrm{sgn}(z)$] in a slab geometry considered before, which is defined by
\begin{align}
\delta\rho(z)=-\frac{e}{L_xL_y}\sum_{k_x,k_y}\sum_{\varepsilon_n\le \varepsilon_F}\left|\psi_{n}(\bm{k}_\parallel, z)\right|^2 - Q,
\label{Charge-def}
\end{align}
where $e>0$ is the elementary charge, $L_i=N_i a$ (with $N_i$ being the number of lattice sites in the $i$ direction and $a$ being the lattice constant), $\psi_{n}(\bm{k}_\parallel, z)$ is the layer-resolved wave function of band $n$ with momentum $\bm{k}_\parallel=(k_x,k_y)$, and we have introduced a charge $Q$ that satisfies the charge conservation law $\int_{-L_z/2}^{L_z/2} dz\, \delta\rho(z)=0$.
We set $\varepsilon_F=0$.
As we have seen in Fig.~\ref{Fig2}, the localized mode emerging at the parallel DW changes its dispersion as a function of $V_0$.
In order to obtain the total charge localized at the DW, we calculate $\delta\rho(z)$ for the case where the dispersion of the localized mode is fully gapped [see Fig.~\ref{Fig2}(d)].
The $z$ dependence of $\delta\rho(z)$ is shown in Fig.~\ref{Fig3}(a).
We see that the localized mode emerging at the parallel DW has a finite charge.
This result can be understood as follows.
As shown in Fig.~\ref{Fig3}(b), in the presence of a uniform mass $V_0$, the surface states have opposite charges to each other, which results in a finite electric polarization $P_z$ that is perpendicular to the nodal line \cite{Ramamurthy2017,Comment4}.
Then, it turns out that the localized mode emerging at the parallel DW has a finite charge $Q_{\mathrm{DW}}$ that is twice the magnitude of the charges at the surfaces: $|Q_{\mathrm{DW}}|=2|P_z|$.
Note that the charge at the parallel DW becomes smaller when its dispersion is gapless [as in Fig.~\ref{Fig2}(c)], because there are unoccupied electron states in its dispersion.
Finally, we find that the localized mode emerging at the perpendicular DW does not have a charge \cite{Comment8}.
%

%%%%%%%%%%%%
{\it Topological characterization of the localized mode.---}Here, we discuss the presence of the localized mode emerging at the parallel DW from a topological consideration on the zero-energy mode \cite{Volovik-book}.
Let us focus on the spectrum around the bulk band extrema $(k_r, k_z)=(\pm k_{\mathrm{NL}}, 0)$ with $k_{\mathrm{NL}}=\sqrt{k_0^2/m_1}$, since the zero-energy mode emerges around them.
Notice that, around the bulk band extrema, the Hamiltonian~(\ref{Hamiltonian-continuum}) can be rewritten as a semiclassical massive Dirac Hamiltonian
\begin{align}
\mathcal{H}(k_\delta, k_z, z)
&\approx -2m_1k_{\mathrm{NL}}k_\delta\sigma_x+v_z k_z\sigma_y+V(z)\sigma_z \nonumber\\
&\equiv\bm{g}(k_\delta,k_z,z)\cdot\bm{\sigma},
\end{align}
where we have defined the momentum $k_\delta\, (\ll 1)$ that satisfies $k_x^2+k_y^2=(k_{\mathrm{NL}}+k_\delta)^2\approx (k_{\mathrm{NL}})^2+2k_{\mathrm{NL}}k_\delta$.
The number of the zero-energy modes is related to the topological charge $N_3$ of the Fermi point located in the semiclassical 3D momentum-real space $(k_\delta, k_z, z)$.
In terms of the semiclassical Green's function $\mathcal{G}=[k_0-\mathcal{H}(k_\delta, k_z, z)]^{-1}$, the topological charge of the 4D Fermi point $(k_0, k_\delta, k_z, z)=(0, 0, 0, 0)$ is given by \cite{Volovik-book}
\begin{align}
N_3\equiv\frac{1}{24\pi^2}\epsilon_{ijkl}\mathrm{tr}\int_{\sigma_3}dS_l\, \mathcal{G}\partial_i\mathcal{G}^{-1}\mathcal{G}\partial_j\mathcal{G}^{-1}\mathcal{G}\partial_k\mathcal{G}^{-1},
\label{topological-charge}
\end{align}
where the integration is done over an arbitrary 3D surface $\sigma_3$ enclosing the Fermi point $(k_0, k_\delta, k_z, z)=(0, 0, 0, 0)$.
Equation~(\ref{topological-charge}) can be evaluated on the infinite planes at $z=\pm z_0$ which face each other across the DW \cite{Comment2}.
Namely, $N_3$ is given by the difference between the topological charges at each side of the DW: $N_3=N_3(z_0)-N_3(-z_0)$ \cite{Volovik-book}.
Here,
\begin{align}
N_3(z_0)&=\frac{1}{4\pi}\int dk_\delta dk_z \frac{1}{|\bm{g}|^3}\bm{g}\cdot[\partial_{k_\delta}\bm{g}\times\partial_{k_z}\bm{g}]\nonumber \\
&=\frac{1}{2}\mathrm{sgn}[V(z_0)],
\label{topological-charge2}
\end{align}
from which we immediately get $N_3=1$.
This is consistent with the number of the localized modes around the bulk band extrema.
Note that the value of $N_3(z_0)$ remains unchanged regardless of the value of $V(z_0)$.
Since the regions with small and large values of $V(z_0)$ are adiabatically connected without the bulk-gap closing, it turns out that the origin of the localized mode is topological even when its dispersion does not cross the zero-energy level.

\begin{figure}[!t]
\centering
\includegraphics[width=\columnwidth]{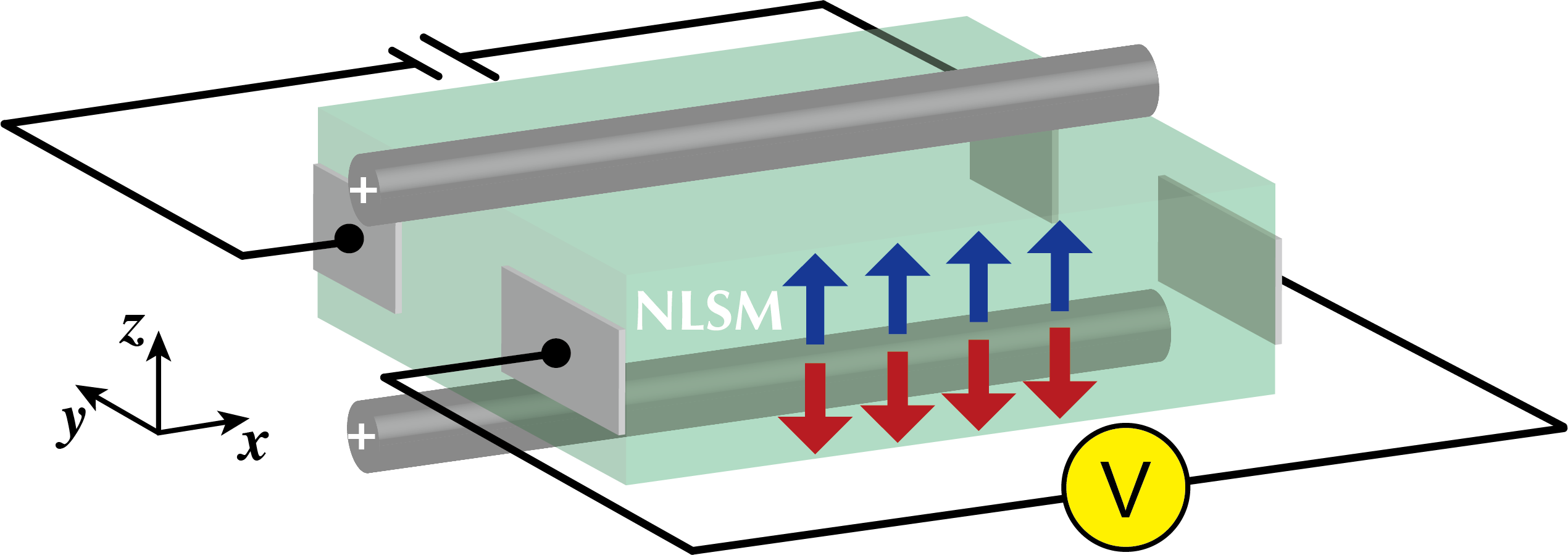}
\caption{A possible experimental setup for realizing the topologically confined localized mode in a $\mathcal{PT}$-symmetric NLSM with the nodal line located on the $x$-$y$ plane.
A weak uniaxial pressure or strain is applied.
Red and blue arrows represent tail-to-tail electric polarizations induced in the bulk.
}
\label{Fig4}
\end{figure}

%%%%%%%%%%%%
{\it Possible experimental realization.---}Let us briefly discuss a possible realization of the topologically confined localized mode.
As we have seen in Fig.~\ref{Fig3}, a mass $V_0$ induces a uniform or spatially-dependent electric polarization along the direction perpendicular to the nodal-line plane.
Inversely, an externally generated electric polarization induces a mass.
Since the original nodal-line system is metallic, a weak uniaxial pressure or strain will need to be applied to induce a small energy gap (i.e., a uniform mass) \cite{Rui2018} before applying an electric voltage for generating the electric polarization.
We show a schematic setup in Fig.~\ref{Fig4}.
Two electrodes of equal sign are attached to the top and bottom surfaces of the thin film of a $\mathcal{PT}$-symmetric NLSM.
The equal sign of the electrodes will realize head-to-head (or tail-to-tail) electric polarizations in the bulk near the two surfaces.
Namely, when the electric fields due to the electrodes are strong enough to compensate the small uniform mass (i.e., uniform electric polarization) induced by the uniaxial pressure or strain, a DW mass $V(z)$ that changes its sign at some point $z_0$ will be realized.
Here, we note that the shape of the electrodes is important in our setup \cite{Comment9}.

The magnitudes of the induced electric polarizations (or equivalently the induced charge at the DW $Q_{\mathrm{DW}}$) are related to the magnitudes of the electric fields due to electrodes.
As we have seen, the charge $Q_{\mathrm{DW}}$ takes the maximum value when the dispersion of the localized mode is gapped.
In other words, its dispersion is gapless when $Q_{\mathrm{DW}}$ is small.
This means that the dispersion of the localized mode is externally tunable, as seen in Fig.~\ref{Fig2}.
Then, the change in the conductivity on the $x$-$y$ plane as a function of the magnitudes of the electric fields due to electrodes will be direct evidence for the observation of the localized mode.

The energy cost for realizing head-to-head or (tail-to-tail) electric polarizations, i.e., a charged DW, can be roughly estimated in terms of the electric field strength $E$ as follows:
\begin{align}
\delta\mathcal{E}=\mathcal{E}_{\mathrm{cDW}}-\mathcal{E}_{\mathrm{uni}}=\frac{1}{2}\epsilon E^2,
\label{Energy-cost}
\end{align}
where $\epsilon$ is the dielectric constant and $\mathcal{E}_{\mathrm{cDW}(\mathrm{uni})}=\frac{1}{L^3}\sum_{k_x,k_y}\sum_{\varepsilon_n\le \varepsilon_F}\varepsilon_{n}(\bm{k}_\parallel)$ with $L^3=L_xL_yL_z$ is the energy density of the lattice model with the electric structure shown in Fig.~\ref{Fig3}(a) [Fig.~\ref{Fig3}(b)].
Taking Ca$_3$P$_2$ \cite{Chan2016} as a candidate material, the energy eigenvalues $\varepsilon_{n}(\bm{k}_\parallel)$ are given in units of 0.1 eV.
Substituting possible values $a\approx 5\ $\AA \ and $\epsilon/\epsilon_0\approx 50$, we obtain $E\sim 10^5\ $V/cm.
We also find the polarization strength in Fig.~\ref{Fig3} as $|P_z|\approx 1.6\ \mathrm{\mu C/cm}^2$, which is an order of magnitude smaller than the typical value ($P_0 \approx 30\ \mathrm{\mu C/cm}^2$) in conventional ferroelectrics \cite{Bednyakov2018}.
Therefore, the required electric field in our study is an order of magnitude larger than that in conventional ferroelectrics \cite{Rani2016} due to the coupling term between polarization and electric field, $-\bm{P}\cdot\bm{E}$, but is still within the experimentally feasible value.

{\it Discussion and summary.---}
Let us compare the present study with previous studies.
The solitonic zero mode in polyacetylene \cite{Jackiw1976,Su1979} and the chiral zero modes in few-layer graphene \cite{Martin2008,Semenoff2008,Yao2009,Jung2011,Ju2015} are the representative examples of the localized modes that emerge at DWs in one and two dimensions, respectively.
It has been understood that these modes emerge as a result of the difference between the bulk topological numbers of the two insulators \cite{Jackiw1976,Su1979,Martin2008,Semenoff2008,Yao2009,Jung2011,Ju2015,Comment6}.
In contrast to this mechanism, our study demonstrates the emergence of a topologically confined state at the interface between two trivial insulators without bulk topological numbers, because the Zak phase (or equivalently the Berry phase in the present case) \cite{Comment5}, which topologically characterizes NLSMs, is no longer quantized in the presence of the $\mathcal{PT}$-breaking mass term in Eq.~(\ref{Hamiltonian-continuum}).
The mechanism in our study is that the 2D subsystem, which can be described by a 2D massive Dirac Hamiltonian, has a topological charge given by Eq.~(\ref{topological-charge2}).

In summary, we have shown the emergence of an in-gap charged localized mode of topological origin in nodal-line semimetals with a gap-opening mass term that forms a domain wall, which realizes a stable, electrically tunable charged domain wall of bulk electric polarizations.
Our study opens up a new direction in possible applications of topological materials in future nanoelectronics.

%%%%%%%%%%%%
\vspace{1.5ex}
\acknowledgements
The authors thank M. Hirayama and A. Yamakage for valuable discussions.
This work was supported by JST CREST Grant No. JPMJCR1874 and No. JPMJCR16F1, and JSPS KAKENHI Grant No. 18H03676 and No. 26103006.
A.S. is supported by the Special Postdoctoral Researcher Program of RIKEN.

%%%%%%%%%%%

\setcounter{figure}{0}
\setcounter{equation}{0}
\setcounter{table}{0}
\renewcommand{\thefigure}{S\arabic{figure}}
\renewcommand{\theequation}{S\arabic{equation}}
\renewcommand{\thetable}{S\Roman{table}}

\begin{widetext}
\vspace{4ex}
\begin{center}
\textbf{{\Large Supplemental Material}}
\end{center}
\section{Lattice Hamiltonian in a slab geometry}
The lattice version of the Hamiltonian~(\ref{Hamiltonian-continuum}) on a cubic lattice (of the lattice constant $a=1$) can be obtained by setting $k^2_i\to 2(1-\cos k_i)$ and $k_i\to \sin k_i$.
In a slab geometry where the system has a finite thickness ($N_z$ layers) in the $z$ direction, we use the identities $e^{ik_za}+e^{-ik_za}=2\cos k_za$ and $e^{ik_za}-e^{-ik_za}=2i\sin k_za$.
The Hamiltonian becomes $H^z_{\mathrm{lattice}}=\sum_{\bm{k}_\parallel}\sum_{n=0}^{N_z-1} H^z_n(\bm{k}_\parallel)$, where
\begin{align}
H^z_n(\bm{k}_\parallel)=
\begin{bmatrix}
c^\dag_{\bm{k}_\parallel,n},\ c^\dag_{\bm{k}_\parallel,n+1}
\end{bmatrix}
\begin{bmatrix}
F^z(\bm{k}_\parallel)\sigma_x+V(n)\sigma_z && m_2\sigma_x-\frac{1}{2}iv_z\sigma_y \\
m_2\sigma_x+\frac{1}{2}iv_z\sigma_y && F^z(\bm{k}_\parallel)\sigma_x+V(n)\sigma_z
\end{bmatrix}
\begin{bmatrix}
c_{\bm{k}_\parallel,n}\\
c_{\bm{k}_\parallel,n+1}
\end{bmatrix},
\label{Hamiltonian-z-dis}
\end{align}
with $F^z(\bm{k}_\parallel)=k_0^2-2m_1(2-\cos k_x -\cos k_y)-2m_2$ and $\bm{k}_\parallel=(k_x,k_y)$.
Similarly, in a slab geometry where the system has a finite thickness ($N_x$ layers) in the $x$ direction, we use the identities $e^{ik_xa}+e^{-ik_xa}=2\cos k_xa$ and $e^{ik_xa}-e^{-ik_xa}=2i\sin k_xa$.
The Hamiltonian becomes $H^x_{\mathrm{lattice}}=\sum_{\bm{k}_\perp}\sum_{n=0}^{N_x-1} H^x_n(\bm{k}_\perp)$, where
\begin{align}
H^x_n(\bm{k}_\perp)=
\begin{bmatrix}
c^\dag_{\bm{k}_\perp,n},\ c^\dag_{\bm{k}_\perp,n+1}
\end{bmatrix}
\begin{bmatrix}
F^x(\bm{k}_\perp)\sigma_x+v_z\sin k_z\sigma_y+V(n)\sigma_z && m_1\sigma_x \\
m_1\sigma_x && F^x(\bm{k}_\perp)\sigma_x+v_z\sin k_z\sigma_y+V(n)\sigma_z
\end{bmatrix}
\begin{bmatrix}
c_{\bm{k}_\perp,n}\\
c_{\bm{k}_\perp,n+1}
\end{bmatrix},
\label{Hamiltonian-x-dis}
\end{align}
with $F^x(\bm{k}_\perp)=k_0^2-2m_1(2-\cos k_y)-2m_2(1-\cos k_z)$ and $\bm{k}_\perp=(k_y,k_z)$.
The energy eigenvalues of the Hamiltonians~(\ref{Hamiltonian-z-dis}) and (\ref{Hamiltonian-x-dis}) with $\bm{k}_\parallel$ and $\bm{k}_\perp$ fixed are obtained respectively by numerically diagonalizing the block-diagonal $2N_z\times 2N_z$ and $2N_x\times 2N_x$ matrices.

\begin{figure}[!b]
\centering
\includegraphics[width=0.8\columnwidth]{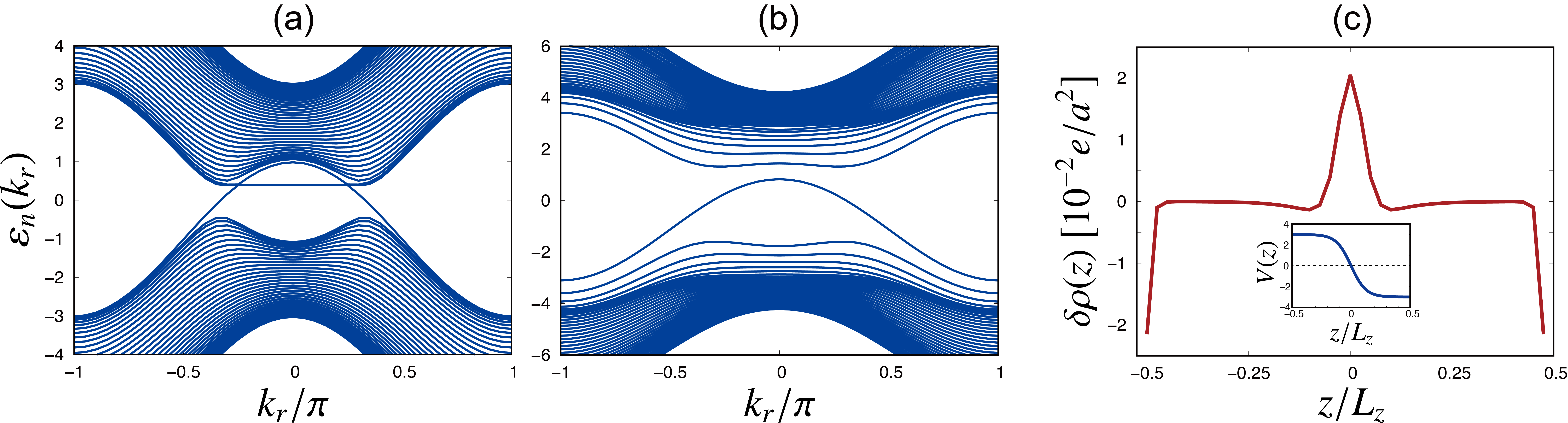}
\caption{Energy spectrum in the presence of a parallel DW $V(z)=V_0\tanh(z/l)$ with (a) $V_0=0.4$ and (b) $V_0=3$.
(c) $z$ dependence of the charge density $\delta\rho(z)$ [Eq.~(\ref{Charge-def})] in the presence of a parallel DW $V(z)=V_0\tanh(z/l)$ with $V_0=3$ and $\varepsilon_F=1$.
In (a)-(c) we set $k_0=m_1=m_2=1$ and $v_z=2$, and $l=N_z/8$.
Note that spectrum and the charge density are inverted, i.e., $\varepsilon_n(k_r)\to -\varepsilon_n(k_r)$ and $\delta\rho(z) \to -\delta\rho(z)$, under the operation $V_0\to -V_0$.
}
\label{FigS1}
\end{figure}
The energy spectrum of the lattice Hamiltonian~(\ref{Hamiltonian-z-dis}) in the presence of the simplest parallel domain wall (DW) structure that takes a step function form [$V(z)=V_0\mathrm{sgn}(z)$] is shown in Fig.~2 of the main text.
Now, let us consider a more realistic domain wall structure in the form $V(z)=V_0\tanh(z/l)$.
We show the energy spectrum of the lattice Hamiltonian~(\ref{Hamiltonian-z-dis}) and the charge density $\delta\rho(z)$ [Eq.~(\ref{Charge-def})] in the presence of a parallel DW $V(z)=V_0\tanh(z/l)$ with $l=N_z/8$ in Fig.~\ref{FigS1}.
We confirm the presence of the in-gap charged localized mode [Figs.~\ref{FigS1}(a) and (b)] and the realization of a charged DW of electric polarizations [Figs.~\ref{FigS1}(c)].
The point is that the emergence of the in-gap charged localized mode does not depend on detailed domain wall structures, because of its topological origin.
Namely, the number of the localize modes is determined by the sign change in the mass term $V(z)$, as is seen in Eq.~(\ref{topological-charge2}).

Next, using the lattice Hamiltonian~(\ref{Hamiltonian-x-dis}), we consider the case of the perpendicular DW, $V(x)=V_0\mathrm{sgn}(x)$.
For reference we show the spectrum for the uniform mass case [i.e., $V(x)=V_0$] in Fig.~\ref{FigS2}(a), from which we see that an energy gap opens.
Note that, unlike the case of the parallel DW, we do not find surface states.
Nonetheless, as shown in Fig.~\ref{FigS2}(b), it is interesting that two in-gap localized modes still emerge at $x=0$ even in the absence of the surface states.
We find that the dispersion of these localized modes never crosses the zero-energy level regardless of the value of the DW mass $V_0$.

The result that the dispersion of the localized modes emerging at the perpendicular DW never crosses the zero-energy level means that the topological consideration we have employed cannot be applied.
This is also understood from that the nodal line (or equivalently Dirac cone) does not exist anymore even without a mass term in the bulk spectrum of the system in a slab geometry with a finite thickness in the $x$ direction, which implies that the origin of those localized modes is not topological.
\begin{figure}[!b]
\centering
\includegraphics[width=0.55\columnwidth]{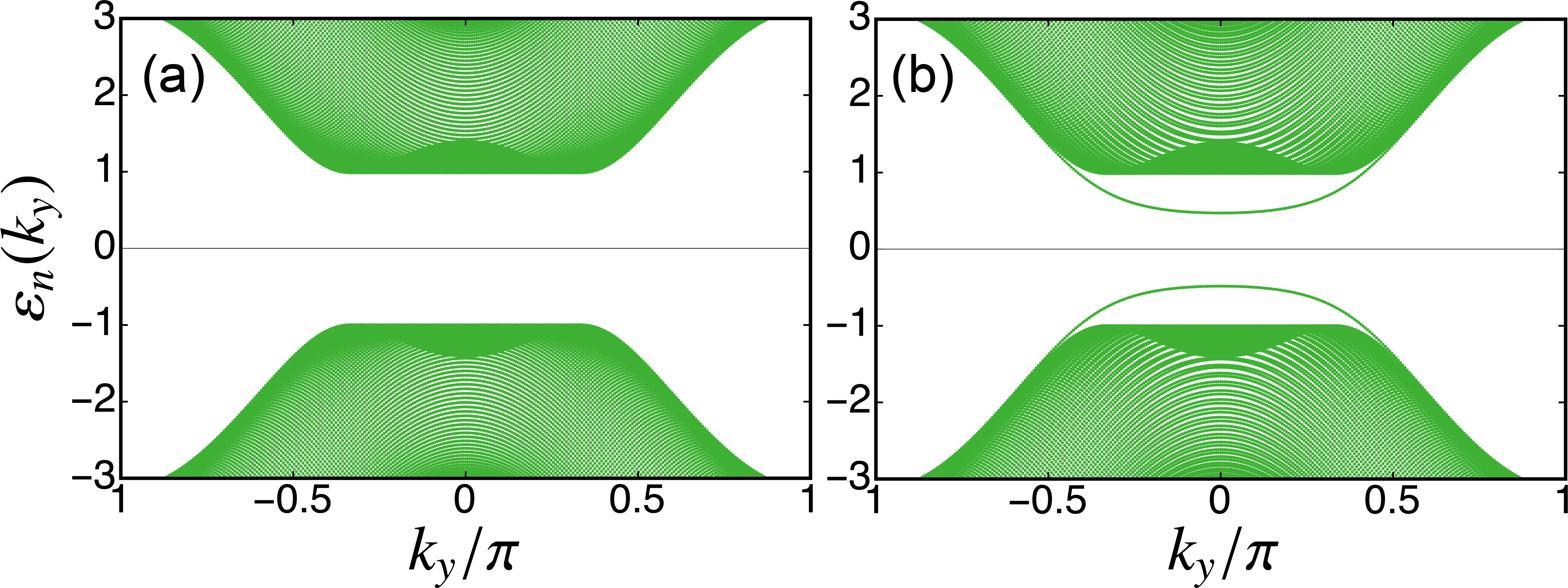}
\caption{Energy spectrum (a) in the presence of a uniform mass with $V_0=1$ and $k_z=0$, and (b) in the presence of a perpendicular DW with $V_0=1$ and $k_z=0$.
In (a) and (b) the parameters are set to $k_0=m_1=m_2=1$ and $v_z=2$.
Note the absence of the surface states.
}
\label{FigS2}
\end{figure}

\section{Semi-analytical solution for the localized modes}
\subsection{The case of the parallel domain wall}
First, we continue our calculation presented in the main text for the case of the parallel DW $V(z)=V_0\mathrm{sgn}(z)$.
The first two boundary conditions are the matching of the wave function and its derivative at $z=0$:
\begin{align}
u^+(0^+)=u^-(0^-)\ \ \ \ \mathrm{and}\ \ \ \ \partial_z u^+(0^+)=\partial_z u^-(0^-).
\label{Boundary-condition1and2}
\end{align}
The third condition is obtained by taking the $0^\pm$ limits in Eq.~(\ref{First-Eq}):
\begin{align}
m_2 \partial_z^2 u^+(0^+)=m_2 \partial_z^2 u^-(0^-)+2V_0u^+(0^+).
\label{Boundary-condition3}
\end{align}
The fourth condition is obtained by first differentiating Eq.~(\ref{First-Eq}) and then taking its $0^\pm$ limits:
\begin{align}
m_2 \partial_z^3 u^+(0^+)=\ m_2 \partial_z^3 u^-(0^-)-2V_0\partial_z u^+(0^+)-v_z[\partial_z^2 u^+(0^+)-\partial_z^2 u^-(0^-)].
\label{Boundary-condition4}
\end{align}
The four boundary conditions~(\ref{Boundary-condition1and2})-(\ref{Boundary-condition4}) can be combined into a $4\times 4$ matrix form:
\begin{align}
\begin{bmatrix}
1 && 1 && -1 && -1\\
-\lambda_1^+ && -\lambda_2^+ && \lambda_1^- && \lambda_2^-\\
(\lambda_1^+)^2-\frac{2V_0}{m_2} && (\lambda_2^+)^2-\frac{2V_0}{m_2} && -(\lambda_1^-)^2 && -(\lambda_2^-)^2\\
-(\lambda_1^+)^3-\frac{2V_0}{m_2}\lambda_1^++\frac{2V_0}{m_2^2}v_z && -(\lambda_2^+)^3-\frac{2V_0}{m_2}\lambda_2^++\frac{2V_0}{m_2^2}v_z && (\lambda_1^-)^3 && (\lambda_2^-)^3
\end{bmatrix}
\begin{bmatrix}
u_1^+\\
u_2^+\\
u_1^-\\
u_2^-
\end{bmatrix}
=0.
\label{Determinant}
\end{align}
We can obtain numerically the solution for the energy eigenvalue $\varepsilon(k_r)$ from the condition such that the determinant of the $4\times 4$ matrix in Eq.~(\ref{Determinant}) is zero, since $\varepsilon(k_r)$ is contained in $\lambda_{1,2}^\pm$.
We show in Fig.~\ref{FigS3} the $V_0$ dependence of $k_r^0$ [Eq.~(\ref{wavenumber-analytical})] and the energy spectrum $\varepsilon(k_r)$ of the localized mode.
\begin{figure}[!b]
\centering
\includegraphics[width=0.6\columnwidth]{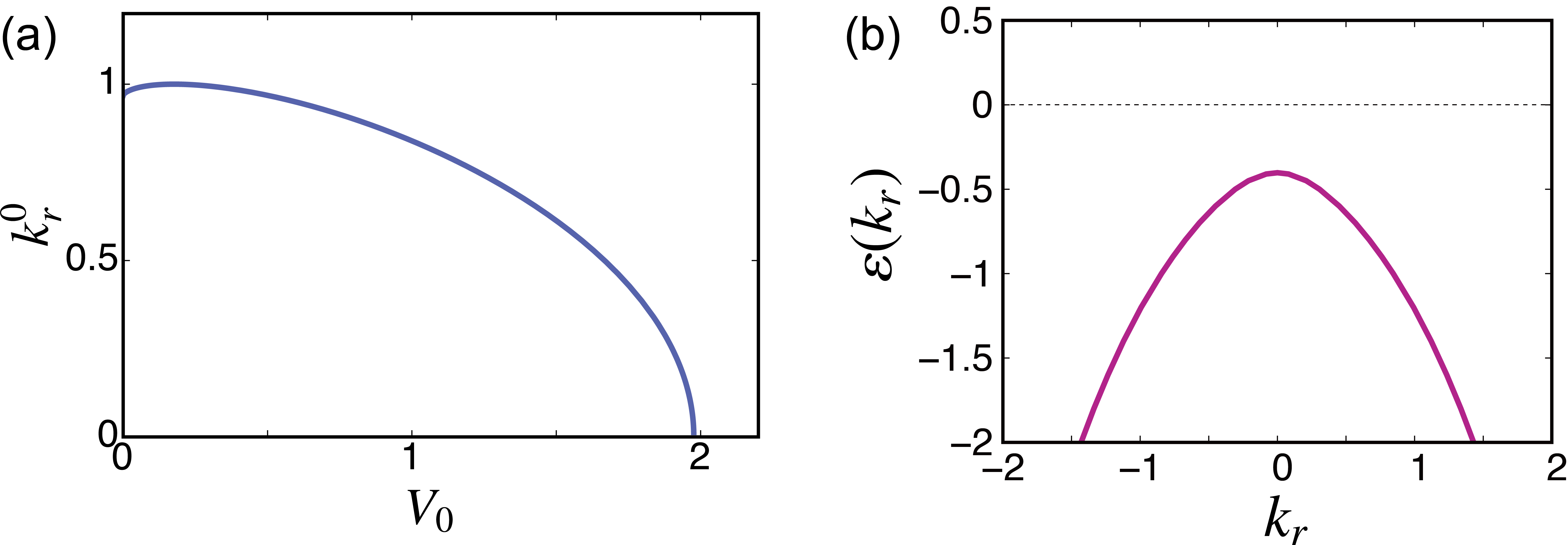}
\caption{(a) Momentum of the zero energy state $k_r^0$ [Eq.~(\ref{wavenumber-analytical})] satisfying $\varepsilon(k_r^0)=0$ as a function of $V_0$.
(b) Energy spectrum $\varepsilon(k_r)$ of the localized mode at $V_0=2.2$.
The parameters are set to $k_0=m_1=m_2=1$ and $v_z=0.5$.
Note that the nodal line is located at $k_r=\pm 1$ when $V_0=0$.
}
\label{FigS3}
\end{figure}

\subsection{The case of the perpendicular domain wall}
Next, we present a calculation for the case of the perpendicular DW $V(x)=V_0\mathrm{sgn}(x)$.
To introduce the spatial dependence in the $x$ direction, we need to set $k_x\to -i\partial_x$ in Eq.~(\ref{Hamiltonian-continuum}).
In this case it is not easy to proceed with the calculation analytically for arbitrary $k_y$ and $k_z$.
Here, let us set $k_z=0$.
Then, the Schroedinger equation $\mathcal{H}(x, k_y, k_z=0)\Psi(x)=\varepsilon\Psi(x)$ with $\Psi(x)=[u(x),v(x)]^T$ reads
\begin{subequations}
\begin{align}
V(x)u+\left[k_0^2-m_1\left(-\partial_x^2+k_y^2\right)\right]v =\varepsilon u,
\label{perp-first-Eq}
\\
\nonumber\\
-V(x)v+\left[k_0^2-m_1\left(-\partial_x^2+k_y^2\right)\right]u=\varepsilon v.
\label{perp-second-Eq}
\end{align}
\end{subequations}
It can be easily checked that, when $V(x)$ has a generic DW structure satisfying $V(-x)=-V(x)$, Eq.~(\ref{perp-second-Eq}) with $v(x)=u(-x)$ or $v(x)=-u(-x)$ is identical to Eq.~(\ref{perp-first-Eq}).
Namely, the solution of the wave functions for the Schroedinger equation has the two forms: $\Psi(x)=[u(x),u(-x)]^T$ or $\Phi(x)=[u(x),-u(-x)]^T$.

Since the wave function $\Psi(x)$ of the mode localized at $x=0$ should be constant in the region where $x$ is far away from zero, we can assume without loss of generality the solution of the form $\Psi(x)\propto e^{-\lambda x}$.
For concreteness let us consider the case of $\Psi(x)=[u(x),u(-x)]^T$.
Then, we have $u(x)=u_0e^{-\lambda x}$ and $v(x)=u(-x)=u_0e^{\lambda x}$.
Substituting these into Eqs.~(\ref{perp-first-Eq}) and (\ref{perp-second-Eq}) and the eliminating $e^{\lambda x}$, we obtain the solution for $\lambda$:
\begin{align}
\lambda=\pm\frac{1}{\sqrt{m_1}}\sqrt{m_1 k_y^2-k_0^2\pm i\sqrt{V_0^2-\varepsilon^2}},
\label{perp-solution-lambda}
\end{align}
for the in-gap states with $|\varepsilon|<V_0$.
Note that the case of $\Phi(x)=[u(x),-u(-x)]^T$ also gives the same solution for $\lambda$ as Eq.~(\ref{perp-solution-lambda}).
We see from Eq.~(\ref{perp-solution-lambda}) that the real part of $\lambda$ can be positive or negative.
Then, the general form of $u(x)$ takes a different form depending on the region with $x>0$ or $x<0$ as follows:
\begin{align}
u^+(x)&=u_1^+e^{-\lambda_1^+ x}+u_2^+e^{-\lambda_2^+ x}\ \ \ (\mathrm{for\ }x>0),\nonumber \\
u^-(x)&=u_1^-e^{-\lambda_1^- x}+u_2^-e^{-\lambda_2^- x}\ \ \ (\mathrm{for\ }x<0),
\end{align}
where the real part of $\lambda_{1,2}^+$ ($\lambda_{1,2}^-$) is positive (negative).
Since there are four unknown constants $u_{1,2}^{\pm}$, we need four boundary conditions for $u^{\pm}(x)$.
The first two boundary conditions are the matching of the wave function and its derivative at $x=0$:
\begin{align}
u^+(0^+)=u^-(0^-)\ \ \ \ \mathrm{and}\ \ \ \ \partial_x u^+(0^+)=\partial_x u^-(0^-).
\label{perp-boundary-condition1and2}
\end{align}
The third condition is obtained by taking the $0^\pm$ limits in Eq.~(\ref{perp-first-Eq}):
\begin{align}
m_1 \partial_x^2 u^+(0^+)=m_1 \partial_x^2 u^-(0^-)+2V_0u^+(0^+).
\label{perp-boundary-condition3}
\end{align}
The fourth condition is obtained by first differentiating Eq.~(\ref{perp-first-Eq}) and then taking its $0^\pm$ limits:
\begin{align}
m_1 \partial_x^3 u^+(0^+)=\ m_1 \partial_x^3 u^-(0^-)-2V_0\partial_x u^+(0^+).
\label{perp-boundary-condition4}
\end{align}
The four boundary conditions~(\ref{perp-boundary-condition1and2})-(\ref{perp-boundary-condition4}) can be combined into the following matrix form:
\begin{align}
\begin{bmatrix}
1 && 1 && -1 && -1\\
-\lambda_1^+ && -\lambda_2^+ && \lambda_1^- && \lambda_2^-\\
(\lambda_1^+)^2-\frac{2V_0}{m_1} && (\lambda_2^+)^2-\frac{2V_0}{m_1} && -(\lambda_1^-)^2 && -(\lambda_2^-)^2\\
-(\lambda_1^+)^3-\frac{2V_0}{m_1}\lambda_1^+ && -(\lambda_2^+)^3-\frac{2V_0}{m_1}\lambda_2^+ && (\lambda_1^-)^3 && (\lambda_2^-)^3
\end{bmatrix}
\begin{bmatrix}
u_1^+\\
u_2^+\\
u_1^-\\
u_2^-
\end{bmatrix}
=0.
\label{perp-determinant}
\end{align}
We can obtain the solution for the energy eigenvalue $\varepsilon(k_y)$ from the condition such that the determinant of the $4\times 4$ matrix in Eq.~(\ref{perp-determinant}) is zero.
The numerically obtained energy eigenvalue $\varepsilon(k_y)$ is shown in Fig.~\ref{FigS4}, which is indeed consistent with the spectrum of the localized mode obtained in a slab system [see Fig.~\ref{FigS2}(b)].
\begin{figure}[!t]
\centering
\includegraphics[width=0.32\columnwidth]{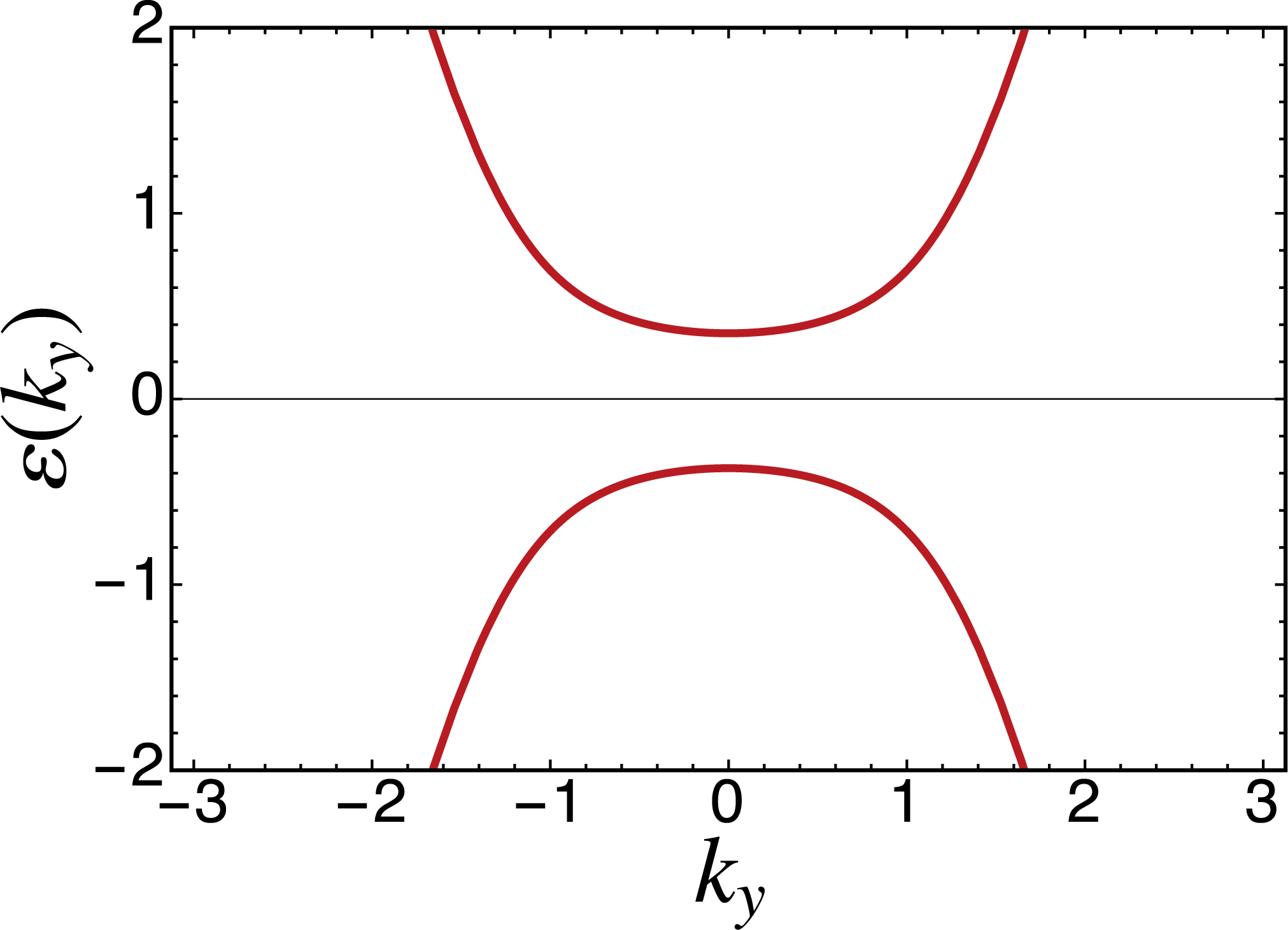}
\caption{The energy spectrum $\varepsilon(k_y)$ with $k_z=0$ of the localized mode at $x=0$.
The parameters are set to $k_0=m_1=V_0=1$.
}
\label{FigS4}
\end{figure}

\section{Possible experimental setup}
Here, we discuss a possible experimental setup for realizing the topologically confined localized mode, focusing on the geometry of the setup.
We consider two cylinder-shaped electrodes of the same sign, as illustrated in Fig.~\ref{FigS5}.
From elementary electromagnetism, it turns out that the electric field from an infinitely long cylinder of radius $R$ is given by
\begin{align}
\vec{E}(r)=\frac{R^2\rho}{2\epsilon r}\vec{e}_r,
\end{align}
where $\rho$ is the charge density of the cylinder, $\epsilon$ is the dielectric constant, $r\ (>R)$ is the distance from the surface of the cylinder, and $\vec{e}_r$ is the unit radius vector.
We define the electric field from the top and bottom electrodes as $\vec{E}_t$ and $\vec{E}_b$, respectively.
Notice that the strength of an electric field is constant in insulators.
Then, the electric fields inside the NLSM is characterized by the angles $\alpha$ and $\beta$, as shown in Fig.~\ref{FigS5}.
We find that the magnitudes of $\vec{E}_t$ and $\vec{E}_b$ inside the NLSM are given respectively by
\begin{align}
E_b(\alpha)=\frac{R\rho}{2\epsilon \sqrt{1+\tan^2\alpha}},\ \ \ \ \ \ \ \ \ 
E_t(\beta)=\frac{R\rho}{2\epsilon \sqrt{1+\tan^2\beta}}.
\label{Setup-electric-fields}
\end{align}
Since $\beta>\alpha$ in the upper half region ($z>0$), the total electric field in the $z$ direction is always positive:
\begin{align}
E_{\mathrm{total}}^z\equiv E_b(\alpha)\cos\alpha - E_t(\beta)\cos\beta >0\ \ \ \ (\mathrm{for}\ z>0).
\label{E_z-upper}
\end{align}
On the other hand, Since $\beta<\alpha$ in the lower half region ($z<0$), the total electric field in the $z$ direction is always negative:
\begin{align}
E_{\mathrm{total}}^z <0\ \ \ \ (\mathrm{for}\ z<0).
\label{E_z-lower}
\end{align}
Eqs.~(\ref{E_z-upper}) and (\ref{E_z-lower}) indicate that head-to-head electric polarizations can be realized in the bulk.

Although we have considered above the simplest setup such that there is no background electric field $\bm{E}_0$ for clarity, it is straightforward to generalize the setup to include $\bm{E}_0$.
Namely, the total electric field in the $z$ direction becomes
\begin{align}
E_{\mathrm{total}}^z= E_0^z + E_b(\alpha)\cos\alpha - E_t(\beta)\cos\beta,
\end{align}
which means that the position $z_0$ where $E_{\mathrm{total}}^z$ changes its sign deviates from $z_0=0$.
\begin{figure}[!t]
\centering
\includegraphics[width=0.5\columnwidth]{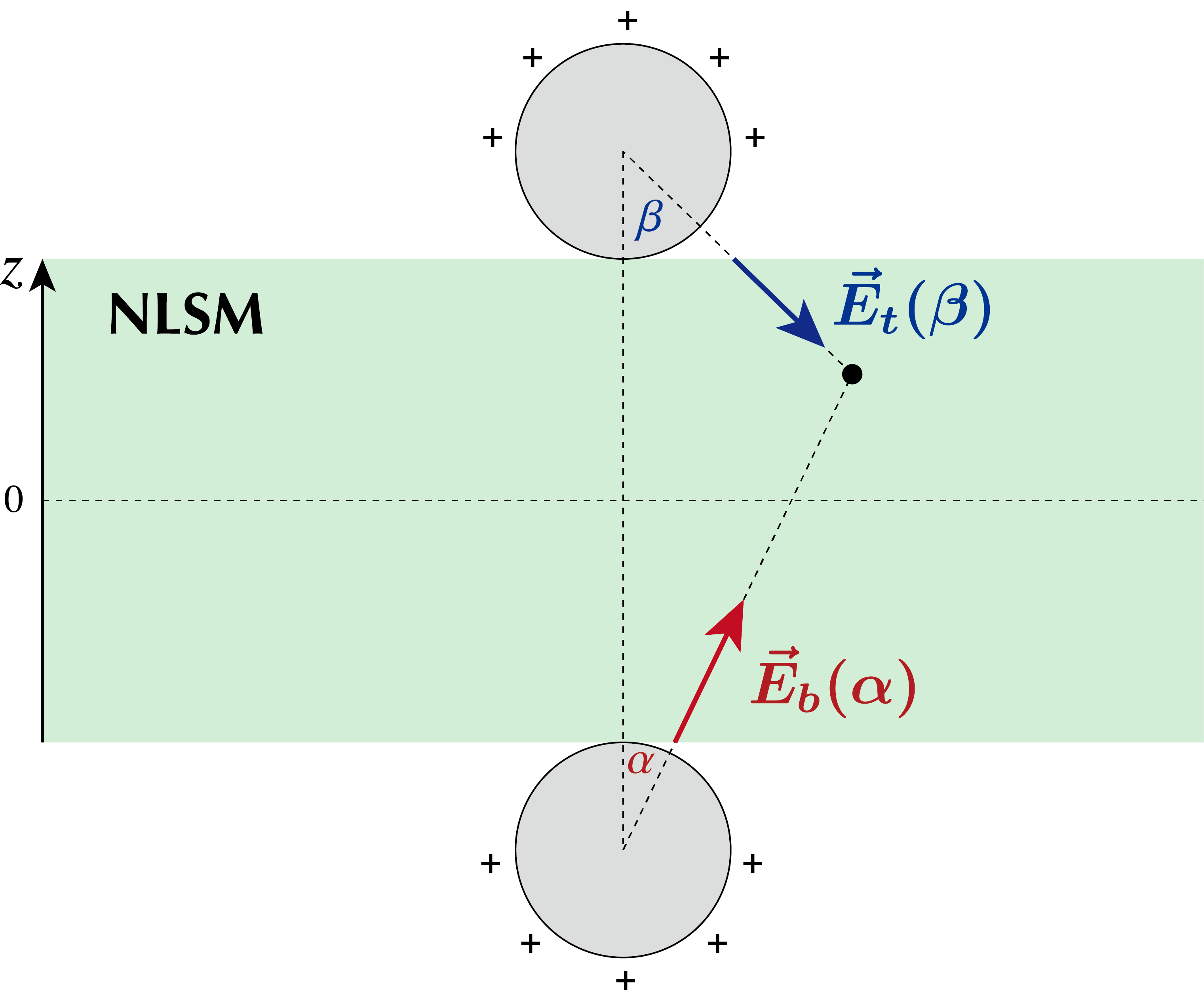}
\caption{Schematic illustration of a possible experimental setup for realizing head-to-head (or tail-to-tail) electric polarizations in the bulk.
The two grey circles indicate long cylinder-shaped electrodes of the same sign.
The two angles $\alpha$ and $\beta$ are defined as the angles from the $z$ axis.
}
\label{FigS5}
\end{figure}

In our experimental setup, the $z$ dependence of the electric field does not take a step function form, but is a smooth function of $z$.
Although it is difficult to simulate the electronic structure in the presence of the electric fields~(\ref{Setup-electric-fields}), we believe that the realization of the charged DW originating from the charged localized mode in our setup is possible.
This is because the origin of the localized mode is topological, i.e., its emergence does not depend on detailed domain wall structures.
Namely, the number of the localize modes is determined by the sign change in the mass term $V(z)$, as is seen in Eq.~(\ref{topological-charge2}).
Actually, we have confirmed in Fig.~\ref{FigS1} the emergence of the charged localized mode in the presence of a parallel DW described by a smooth function of the form $V(z)=V_0\tanh(z/l)$ with $l=N_z/8$.

We also note that the electric field that is perpendicular to the $z$ direction does not cause serious problems.
Recall that the side surfaces that are perpendicular to the nodal line do not have conductive surface states, as can be seen from Fig.~\ref{FigS2}.
Then, although the localized charge may accumulate near the side surfaces, the localized charge never spreads over the side surfaces.
In other words, the head-to-head (or tail-to-tail) electric polarizations may tilt (i.e., have both $z$ and $y$ components), but even in this case the DW due to the $z$ component of the electric polarizations is present.

\end{widetext}


\begin{thebibliography}{999}
\bibitem{Comment7}
In the present case, the spatial-inversion and time-reversal operators are given by $\mathcal{P}=\sigma_x$ and $\mathcal{T}=\mathcal{K}$ (with $\mathcal{K}$ the complex conjugation operator), respectively.
Then, the combined $\mathcal{PT}$ operator is given by $\mathcal{PT}=\sigma_x\mathcal{K}$.
In order to break $\mathcal{PT}$ symmetry, either $\mathcal{P}$ or $\mathcal{T}$ symmetry needs to be broken.
Because $\sigma_z$ preserves time-reversal symmetry but breaks spatial-inversion symmetry, it turns out that spatial-inversion symmetry breaking perturbation is required to induce the gap-opening term $V\sigma_z$.
\bibitem{Comment5}
The Berry phase along a loop $C$  enclosing the nodal line is defined as $\nu=\sum_{n}^{\mathrm{occupied}}\oint_C d\bm{k}\cdot\bm{\mathcal{A}}^n(\bm{k})$, where $\mathcal{A}_a^n(\bm{k})=i\langle u_{\bm{k}}^n|\partial_{k_a}| u_{\bm{k}}^n\rangle$ ($a=x,y,z$) is the Berry connection with $| u_{\bm{k}}^n\rangle$ the Bloch wave function of band $n$.
On the other hand, the Zak phase is defined as $\theta(k_x, k_y)=\sum_{n}^{\mathrm{occupied}}\int dk_z \mathcal{A}_z^n(\bm{k})$.
From the Hamiltonian~(\ref{Hamiltonian-continuum}) the Berry connection is obtained as $\mathcal{A}_z^\pm(\bm{k})=\pm v_z V[(k_0^2-m_1k_r^2-m_2k_z^2)\mp E_{\bm{k}}]/[2E_{\bm{k}}(v_z^2k_z^2+V^2)]$ and $\mathcal{A}_x^n(\bm{k})=\mathcal{A}_y^n(\bm{k})=0$, where $+$($-$) indicates the conduction (valence) band.
Since $\int dk_z \mathcal{A}_z^n(\bm{k})$ is $\pi$ (mod $2\pi$) for a given set of $(k_x, k_y)$ located inside the nodal line and otherwise it is zero, it turns out that $\nu=\theta$ in the present case.
\bibitem{Comment3}
Note that the case of $\Phi(z)=[u(z),-u(-z)]^T$ also gives the same solution for $\lambda$ as Eq.~(\ref{Solution-lambda}).
\bibitem{Comment4}
Note that the polarization $P_z$ is not quantized due to the inversion symmetry breaking mass $V_0$.
In the limit of $V_0\to 0$ and $N_i\to \infty$ ($i=x,y,z$), we reproduce numerically an expression $P_z=\mathrm{sgn}(V_0)(em/8\pi)$ for small values of $m=k_0^2/m_1<1$, which was obtained in Ref.~\cite{Ramamurthy2017}.
Because $\sqrt{m}$ is the radius of the nodal line, this expression is rewritten as $P_z=\mathrm{sgn}(V_0)(A_{\mathrm{NL}}/4\pi^2)(e/2)$ with $A_{\mathrm{NL}}=\pi m$ being the area surrounded by the nodal line.
\bibitem{Comment8}
This may be understood from that there are no surface states in the lattice model with a uniform mass in a slab geometry where the thickness is finite in the $x$ direction \cite{Comment-SM}.
\bibitem{Comment2}
When the distance between the infinite planes ($=2z_0$) is sufficiently small, the contribution from the fluxes that do not intersect the planes can be neglected.
\end{thebibliography}
\end{document}